\documentstyle{article}

\def\abstract#1{\vskip 7mm 
        \begin{center}{\large Abstract}\par \smallskip
                \begin{minipage}[c]{12cm}
                        \small #1
                \end{minipage}
        \end{center}
}
\def\title#1{\begin{center}{\Large\bf #1}\end{center}}
\def\author#1{\vskip 5mm \begin{center}{#1}\end{center}}
\def\address#1{\begin{center}{\it #1}\end{center}}

\begin{document}

\newcommand{\sikib}{\begin{eqnarray}}
\newcommand{\sikie}{\end{eqnarray}}
\newcommand{\maru}{\partial}
\newcommand{\gb}{\bar{g}}
\newcommand{\La}{\Lambda}
\newcommand{\la}{\lambda}
\newcommand{\al}{\alpha}
\newcommand{\be}{\beta}
\newcommand{\Om}{\Omega}
\newcommand{\om}{\omega}
\newcommand{\Mb}{\bar{M}}
\newcommand{\Jb}{\bar{J}}
\newcommand{\lb}{\bar{l}}
\newcommand{\Sb}{\bar{S}}


\title{BTZ Black Hole Entropy in Higher Curvature Gravity
}
\author{Hiromi SAIDA
\footnote{E-mail: saida@phys.h.kyoto-u.ac.jp}
}
\address{Graduate School of Human and Environmental Studies, 
          Kyoto University, \\
  Sakyo-ku, Kyoto 606--8501, Japan
}

\author{Jiro SODA
\footnote{E-mail: jiro@phys.h.kyoto-u.ac.jp}
}
\address{Department of Fundamental Sciences, FIHS,
          Kyoto University, \\
  Sakyo-ku, Kyoto 606--8501, Japan
}


\abstract{
For the BTZ black hole in the Einstein gravity, a statistical 
entropy has been calculated to be equal to the Bekenstein-Hawking 
entropy. In this paper, its statistical entropy in the higher 
curvature gravity is calculated and shown to be equal to the one 
derived by using the Noether charge method. This suggests that the 
equivalence of the geometrical and statistical entropies of the 
black hole is retained in the general diffeomorphism invariant 
theories of gravity. A relation between the cosmic censorship 
conjecture and the unitarity of the conformal field theory on the 
boundary of $AdS_3$ is also discussed. 
}


\section{Introduction}\label{sec-intro}

The profound understanding of the black hole thermodynamics is a 
clue for approaching the quantum gravity. Although the final form of 
the quantum gravity is covered with a veil of mystery, it would be 
legitimate to regard the diffeomorphism invariance as the key to the 
mystery. Hence the black hole thermodynamics in the diffeomorphism 
invariant theories of gravity should be studied to obtain the 
insights into the quantum gravity. The first law of black holes in 
such theories of gravity has already been established \cite{ref-IW1} 
\cite{ref-JKM1}. The zeroth and second laws, especially in the 
higher curvature gravity, have been investigated, for example in 
ref. \cite{ref-JKM2}. All the black hole entropies treated in the 
above works are of the integral of geometrical quantities on a 
spatial section of the event horizon. The geometrical entropy in an 
N-dimensional higher curvature gravity can be calculated by using 
the Noether charge method \cite{ref-IW1} 
\cite{ref-JKM1} to give
  \sikib
    S_{IW} =
      -\frac{1}{8G} \oint_{H} dx^{N-2} \sqrt{h} \,
      \frac{ \maru f }{ \maru R_{\mu\nu\al\be} } \,
      \epsilon_{\mu\nu}\epsilon_{\al\be} \, ,
  \label{eq-I1}
  \sikie
where $H$ is the spatial section of the event horizon, $h$ is the 
determinant of the induced metric on $H$, $f$ is the higher 
curvature Lagrangian and $\epsilon_{\mu\nu}$ is the binormal to 
$H$ \cite{ref-JKM1}. When we denote $f = R + $(higher curvature 
terms), the first term of this Lagrangian contributes to the entropy 
(\ref{eq-I1}) as the Bekenstein-Hawking term $A/4G$, where $A$ is 
the area of $H$. The geometrical expression of the black hole 
entropy obtained in ref. \cite{ref-IW1} is consistent with the well 
known results previously obtained for the Einstein gravity, however 
it does not reveal statistical origin of the entropy. Statistical 
explanation of the entropy is remained as an open question. 

For the Einstein gravity, a statistical derivation of the 
Bekenstein-Hawking entropy has been carried out \cite{ref-BH1} 
\cite{ref-S1} for a black hole in three dimensional anti-de Sitter 
spacetime ($AdS_3$) which is called the BTZ black hole 
\cite{ref-BTZ1} \cite{ref-BTZ2} \cite{ref-C1}. Although several 
issues remain open to be resolved \cite{ref-C2}, this is one of the 
important examples which demonstrate the equality between the 
Bekenstein-Hawking entropy and a statistical entropy. It is our 
central interest to give a statistical interpretation of the black 
hole entropy in the diffeomorphism invariant theories of gravity. In 
this paper, we attempt to derive the statistical entropy of the BTZ 
black hole in the higher curvature gravity, and the computation of 
it in refs. \cite{ref-BH1} and \cite{ref-S1} is modified to be 
available for this case \cite{ref-SS1}.

In section \ref{sec-EG} we review the statistical derivation of the 
BTZ black hole entropy in the Einstein gravity. Section 
\ref{sec-HCG} is devoted to the computation of the entropy in the 
higher curvature gravity, and our results are illustrated with a 
concrete example. Finally we give summary and discussion in section 
\ref{sec-SD}. Throughout this paper we use the unit, $c=\hbar=1$.


\section{BTZ Black Hole Entropy in the Einstein Gravity}
\label{sec-EG}

For the preparation of our purpose, in this section, we review the 
BTZ black hole \cite{ref-BTZ1} \cite{ref-BTZ2} \cite{ref-C1} and its 
statistical entropy in the Einstein gravity \cite{ref-S1}. The 
action we treat here is
  \sikib
    I = \frac{1}{16\pi G} \int d^3x \, 
       (R - 2\La) \sqrt{-g} + B \, ,
  \label{eq-EG1}
  \sikie
where $\La$ is the cosmological constant and $B$ is a surface term 
which is needed to let the variation of $I$ make sense 
\cite{ref-BTZ1}.

\subsection{BTZ Black Hole}

Because three dimensional gravity have no dynamical degrees of 
freedom, the BTZ black hole spacetime is locally equivalent to 
$AdS_3$. Then it is enough for our purpose to consider only a 
constant negative curvature spacetime. The Riemann tensor of the 
constant curvature spacetime is expressed as
  \sikib
    R_{\mu \nu \al \be}
      = \frac{R}{6}
        (g_{\mu \al} g_{\nu \be} - g_{\mu \nu}g_{\al \be}) \, ,
  \label{eq-EG2}
  \sikie
where $R$ is the Ricci scalar which is related to the cosmological 
constant through the Einstein equation: $R = 6\La$. Further we can 
set $R = -6/l^2$ where $l$ is the curvature radius of $AdS_3$. 
Note that $l$ is related to the cosmological constant, $\La=-6/l^2$. 
The BTZ black hole metric is \cite{ref-BTZ1}
  \sikib
    ds^2 = -N^2 dt^2 + \frac{1}{N^2} dr^2
          + r^2 \left[ d\phi + N^{\phi} dt \right]^2 \, ,
  \label{eq-EG3}
  \sikie
where $\phi \sim \phi + 2\pi$, $N^2 = (r/l)^2 +(4GJ/r)^2 -GM/l$, 
$N^{\phi} = -4GJ/r^2$, and $M$ and $J$ are the mass and the angular 
momentum of the black hole, respectively. The event horizon is 
at $r_+ = \sqrt{2Gl(M+J)} + \sqrt{2Gl(M-J)}$, further the 
Bekenstein-Hawking entropy is calculated to be 
  \sikib
    S_{BH} = \frac{1}{4G}\oint_{r_+} d\phi \sqrt{ g_{\phi \phi} }
           = \frac{\pi}{4G}
             \left[ \, \sqrt{ 8Gl(M+J) }
                + \sqrt{ 8Gl(M-J) } \, \right] \, .
  \label{eq-EG5}
  \sikie

\subsection{Statistical Entropy}

The BTZ black hole entropy can be calculated by counting the number 
of states \cite{ref-S1} \cite{ref-C2}. The counted states are in the 
Hilbert space operated by the quantum counterparts (generators) of 
asymptotic symmetry of asymptotically $AdS_3$, whose definitions are 
described below \cite{ref-BH1}. The asymptotically $AdS_3$ metrics 
are defined as: 
  \sikib
    ds^2 &=& \left[ -\frac{r^2}{l^2} + O(1) \right] dt^2
         + \left[ \frac{l^2}{r^2}
                + O\left( \frac{1}{r^4} \right)
           \right] dr^2
         + \left[ r^2 + O(1) \right] d\phi^2 \nonumber \\
    && \quad
         + O\left( \frac{1}{r^3} \right) dt dr
         + O(1) dt d\phi
         + O\left( \frac{1}{r^3} \right) dr d\phi \, ,
  \label{eq-EG13}
  \sikie
where $0<\phi \leq 2\pi$. The global $AdS_3$, which denotes the 
$AdS_3$ spacetime without any source in it, is included in these 
metrics. It is obvious that the BTZ black hole is of asymptotically 
$AdS_3$. The asymptotic symmetry of asymptotically $AdS_3$ is 
described by the following Killing vectors which preserve the 
boundary condition of the metric \cite{ref-BH1} \cite{ref-S1}
  \sikib
    \xi^t &=& l(T^+ + T^-)
           + \frac{l^3}{2r^2}(\maru_+^2 T^+ + \maru_-^2 T^-)
           + O\left( \frac{1}{r^4} \right)
      \, , \nonumber \\
    \xi^r &=& -r(\maru_+ T^+ + \maru_- T^-)
          + O\left( \frac{1}{r} \right)
      \, , \label{eq-EG14} \\
    \xi^{\phi} &=& T^+ - T^-
           - \frac{l^2}{2r^2}(\maru_+^2 T^+ - \maru_-^2 T^-)
           + O\left( \frac{1}{r^4} \right)
      \, , \nonumber
  \sikie
where $T^{\pm}=T^{\pm}(x^{\pm})$ and $x^{\pm}=t/l\pm\phi$. The 
Killing vectors in this form include those of the global $AdS_3$. 

The algebra satisfied by the quantum counterparts (generators) of 
asymptotic symmetry of asymptotically $AdS_3$ should be of the 
quantization of the classical algebra satisfied by classical 
generators of the symmetry, which is just the Hamiltonian generating 
the diffeomorphism along the Killing vector $\xi $\cite{ref-BH1}. 
Such a Hamiltonian is expressed as 
$H[\xi] = \int_{\Sigma} d^2 x \, \xi^a {\cal H}_a + Q[\xi]$, where 
$\Sigma$ is the spatial hypersurface, $\xi^a$ ($a = \bot, 1, 2$) 
are the vertical and parallel components of $\xi$ with respect to 
$\Sigma$, ${\cal H}_a$ are the constraints and $Q[\xi]$ is the 
surface term which is needed to cancel the surface term in the 
variation $\delta H[\xi]$, that is, there is a freedom of additional 
constant to $Q$. Given two Hamiltonians, $H[\xi]$ and $H[\eta]$, the 
algebra of them is obtained by evaluating the Poisson bracket. 
It is shown in ref. \cite{ref-BH1} that such an algebra is the 
central extension of the commutation relation (Lie bracket) of $\xi$ 
and $\eta$, 
$\{ H[\xi], H[\eta] \}_P = H[ \, \{ \xi,\eta \} \, ] + K[\xi,\eta]$, 
where $K[\xi,\eta]$ is the central extension, $\{ \, \}_P$ is the 
Poisson bracket and $\{ \xi,\eta \}$ is the Lie bracket. Further, 
because the constraints ${\cal H}_a$ is weakly zero, it reduces to 
the algebra: 
$\{ Q[\xi], Q[\eta] \}_D = Q[ \, \{ \xi,\eta \} \, ] + K[\xi,\eta]$, 
where $\{ \, \}_D$ is the Dirac bracket. The central extension 
$K[\xi,\eta]$ can be calculated by setting the surface term $Q$ for 
the global $AdS_3$ be zero\cite{ref-BH1}:
  \sikib
    K[\xi,\eta] = \lim_{r\to\infty} \frac{1}{16\pi G} \oint dS_{l}
       \{ \, \widehat{G}^{ijkl} \,
              [ \, \xi^{\bot} g_{ij|k}
              - \xi^{\bot}_{,k} \, (g_{ij} - \hat{g}_{ij})\, ]
          + 2 \, (\, \xi^i + N^i \xi^{\bot} \, ) \, \pi_i^l \,
       \} \, ,
  \label{eq-EG18}
  \sikie
where the latin indices denote the spatial coordinates, 
$\hat{g}_{\mu \nu}$ is the global $AdS_3$, $g_{\mu \nu}$ is the 
asymptotically $AdS_3$ of the form (\ref{eq-EG13}), 
$\widehat{G}^{ijkl} = (1/2)\sqrt{\det(\hat{g}_{ij})} \,
  ( \hat{g}^{ik}\hat{g}^{jl} + \hat{g}^{il}\hat{g}^{jk}
   - 2\hat{g}^{ij}\hat{g}^{kl} )$, 
$dS_l$ is the line element of $r=const.$ circle on the spatial 
hypersurface $\Sigma$, $\pi_i^{\;l} = g^{kl}\pi_{ik}$, 
$\pi_{ij}$ is the conjugate momentum of $g_{ij}$, 
$A_{|k}$ is covariant derivative of quantity $A$ with respect to 
spatial metric $\hat{g}_{ij}$, $\xi^{\bot}= N^{\bot} \xi^{t}$, 
and $N^{\bot}$ and $N^{i}$ are the lapse function and the shift 
vector of $g_{\mu\nu}$ respectively. Here the variation of metric 
from $\hat{g}_{\mu\nu}$ to $g_{\mu\nu}$ is given by 
$\delta g_{\mu \nu} \equiv g_{\mu \nu} - \hat{g}_{\mu \nu}
  = {\cal L}_{\eta} \hat{g}_{\mu \nu}$. 

When we define the Fourier components of $\xi$ as 
$\xi^{\pm}_m \equiv (i/2) \, \xi$ with $T^{\pm} = \exp(imx^{\pm})$ 
($m = 0, \pm 1, \pm 2 \cdots$), the Lie brackets of the Killing 
vectors are calculated to be the Virasoro algebra without central 
extension: 
$\{ \xi^{\pm}_m, \xi^{\pm}_n \} = (m-n) \, \xi^{\pm}_{m+n}$, 
$\{ \xi^{+}_m, \xi^{-}_n \} = O(1/r)$. Then the classical algebra we 
are seeking can be expressed as the Virasoro algebra with the 
central extension given by eq. (\ref{eq-EG18}). That is, the algebra 
of the quantum generators is also the same algebra, which is 
obtained by quantizing the commutation relation through the 
replacement, $\{ \, \}_D \to -i [ \, \, ]$, where $[ \, \, ]$ is the 
commutation relation of quantum operator \cite{ref-BH1}:
  \sikib
    \mbox{ [ } L_m , L_n \mbox{ ] } &=&
         (m-n) \, L_{m+n}
       + \frac{C}{12} \, m(m^2-1) \, \delta_{m+n,0}
      \, , \nonumber \\
    \mbox{ [ } \tilde{L}_m , \tilde{L}_n \mbox{ ] } &=&
         (m-n) \, \tilde{L}_{m+n}
       + \frac{C}{12} \, m(m^2-1) \, \delta_{m+n,0}
      \, , \label{eq-EG17} \\
    \mbox{ [ } L_m , \tilde{L}_n \mbox{ ] } &=& 0
      \, . \nonumber
  \sikie
Here $L_m$ and $\tilde{L}_m$ denote the quantum generators 
corresponding to $\xi^+_m$ and $\xi^-_m$ respectively, and the 
central charge $C$ can be obtained by substituting the metric 
given by eq. (\ref{eq-EG13}) into $g_{\mu\nu}$ in eq. 
(\ref{eq-EG18}) to be a positive constant \cite{ref-BH1},
  \sikib
    C = \frac{3 \, l}{2 \, G} \, .
  \label{eq-EG20}
  \sikie
Thus the quantum gravity of $AdS_3$ is related to the conformal 
field theory (CFT) on the boundary of the spacetime. Such a 
correspondence between the $AdS$ and the CFT is called the AdS/CFT 
correspondence \cite{ref-CC} \cite{ref-NO1}.

With above preparations, we can proceed to a calculation of 
statistical entropy. The calculation follows the ordinary state 
counting of the conformal field theory, where the counted states 
are the eigen states of $L_0$ and $\tilde{L}_0$. For the case that 
the eigen values are large (semi-classical), we can obtain the 
number of states by Cardy's formula \cite{ref-C2} \cite{ref-Ca1} to 
give the statistical entropy:
  \sikib
    S_C \approx 2\pi \sqrt{ \frac{C \, \la}{6} }
              + 2\pi \sqrt{ \frac{C \, \tilde{\la} }{6} } \, ,
  \label{eq-EG21}
  \sikie
where $\la$ and $\tilde{\la}$ are the (large) eigen values of $L_0$ 
and $\tilde{L}_0$ respectively.

Note that $\xi^+_0 + \xi^-_0 = 2l \, \maru_t + O(1/r^4)$ and 
$\xi^+_0 - \xi^-_0 = 2 \, \maru_{\phi} + O(1/r^4)$, then a mass 
operator ${\cal M}$ and an angular momentum operator ${\cal J}$ are 
defined by ${\cal M} \equiv L_0 + \tilde{L}_0$ and 
${\cal J} \equiv L_0 - \tilde{L}_0$, respectively \cite{ref-S1}. For 
the BTZ black hole, we define a quantum state $\left| MJ \right>$ by 
${\cal M} \left| MJ \right> = M \left| MJ \right>$ and 
${\cal J} \left| MJ \right> = J \left| MJ \right>$, further the 
ground state by ${\cal M} \left| 00 \right> = 0$ and 
${\cal J} \left| 00 \right> = 0$, that is, $M=J=0$. The spacetime of 
the ground state asymptotes to the $AdS_3$ as $r\to\infty$, thus 
this ground state can correspond to $\hat{g}_{\mu\nu}$ in eq. 
(\ref{eq-EG18}). Consequently, the above calculations of the central 
charge and the statistical entropy are available for the BTZ black 
hole \cite{ref-C2}. 
The state, $\left| MJ \right>$, provides the eigen values of 
$L_0$ and $\tilde{L}_0$ to be
  \sikib
    \la = \frac{1}{2}(M+J) \quad , \quad
    \tilde{\la} = \frac{1}{2}(M-J) \, ,
  \label{eq-EG25}
  \sikie
then the statistical entropy for the case of large $M$ and $J$ 
becomes
  \sikib
    S_C = \frac{\pi}{4G}
          \left[ \, \sqrt{ 8Gl(M+J) }
                + \sqrt{ 8Gl(M-J) } \, \right] \, .
  \label{eq-EG26}
  \sikie
This coincides with the Bekenstein-Hawking entropy $S_{BH}$ given by 
eq. (\ref{eq-EG5}).


\section{BTZ Black Hole Entropy in the higher Curvature Gravity}
\label{sec-HCG}

In this section, we calculate the statistical entropy of the BTZ 
black hole in the higher curvature gravity. The action we treat here 
is of the form:
  \sikib
    I = \frac{1}{16\pi G} \int d^3 x \,
        f(R_{\mu \nu}, g^{\mu \nu}) \sqrt{-g} + B \, ,
  \label{eq-FT1}
  \sikie
where $f(R_{\mu \nu}, g^{\mu \nu})$ is the Lagrangian of general 
higher curvature gravity and $B$ is the surface term needed by the 
same reason as the action (\ref{eq-EG1}). This action is the most 
generic form of the three dimensional higher curvature gravity 
because the Weyl tensor vanishes in three dimensional spacetime, 
that is, the Riemann tensor can be expressed generally by the Ricci 
tensor, the Ricci scalar and the metric: 
$ R_{\mu\nu\al\be} =
        g_{\mu\al}R_{\nu\be} +g_{\nu\be}R_{\mu\al}
      - g_{\nu\al}R_{\mu\be} -g_{\mu\be}R_{\nu\al} 
      - (1/2) \, ( g_{\mu\al}g_{\nu\be}
                       - g_{\mu\be}g_{\nu\al} )\, R
$. 

When we are interested in the statistical entropy of the BTZ black 
hole in the higher curvature gravity, the entropy is given by the 
same formula as eq. (\ref{eq-EG21}) with the central charge and the 
eigen values of the Virasoro generators in the higher curvature 
gravity. The eigen values (\ref{eq-EG25}) can be easily obtained 
through the Noether charge form of the mass and the angular momentum 
constructed in ref. \cite{ref-IW1}. However, as reviewed in the 
previous section, the central charge is necessary to calculate the 
surface terms of the Hamiltonian, which should be corrected by the 
higher curvature terms in the Lagrangian to induce additional terms 
upon the central charge, eq. (\ref{eq-EG18}). The derivation and 
resultant form of the additional terms will be very complicated. To 
avoid this intricacy, we define new metric and transform the 
original higher curvature frame into the Einstein frame.

\subsection{Frame Transformation}\label{sec-FT}

In this subsection, we review the transformation from the original 
higher curvature frame to the Einstein frame in preparation for 
calculating the central charge. The Euler-Lagrange equation in the 
original frame is obtained as
  \sikib
    \frac{\maru f}{\maru g^{\mu\nu}} - \frac{1}{2} f g_{\mu\nu} =
      \frac{1}{2} \left[
        P_{\al\mu;\nu}^{\quad \;\; ;\al}
      + P_{\al\nu;\mu}^{\quad \;\; ;\al}
      - \Box P_{\mu\nu}
      - g_{\mu\nu} P_{\al\be}^{\quad ;\al\be} \right] \, ,
  \label{eq-FT2}
  \sikie
where 
$P_{\mu\nu} \equiv g_{\mu\al}g_{\nu\be}(\maru f/\maru R_{\al\be})$. 
This is obviously a fourth order differential equation of 
$g^{\mu\nu}$. The new metric which turns out to be that in the 
Einstein frame is defined by \cite{ref-MFF1} \cite{ref-MFF2}
  \sikib
    \gb^{\mu\nu} \equiv
      \left[
        -\det \left( \frac{\maru (f \sqrt{-g}) }{\maru R_{\al\be}}
      \right) \right]^{-1}
      \frac{\maru (f \sqrt{-g}) }{\maru R_{\mu\nu}} \, .
  \label{eq-FT3}
  \sikie
We assume here that the relation (\ref{eq-FT3}) can be inverted to 
give $R_{\mu\nu}$ as a function of $\gb^{\al\be}$ and $g^{\al\be}$ 
(including no derivative of them): 
$ R_{\mu\nu} = R_{\mu\nu}(\gb^{\al\be},g^{\al\be}) $.

The alternative action which is equivalent to the original action 
$I$ and describes the Einstein frame can be defined through the 
Legendre transformation as \cite{ref-MFF1} \cite{ref-MFF2}
  \sikib
    \bar{I} \equiv
      \frac{1}{16\pi G} \int d^3 x \, \sqrt{-\gb} \,
    \left[ \,
       \gb^{\mu\nu}
        R_{\mu\nu}(g^{\al\be}, \maru_{\om} g^{\al\be},
                   \maru_{\tau} \maru_{\om} g^{\al\be})
     - \gb^{\mu\nu} R_{\mu\nu}(\gb^{\al\be},g^{\al\be})
    \right. \nonumber \\
    \left.
     + \frac{ \sqrt{-g} }{ \sqrt{-\gb} } \,
          f(R_{\mu\nu}(\gb^{\al\be},g^{\al\be}),g^{\mu\nu}) \,
      \right] + \bar{B} \, .
  \label{eq-FT5}
  \sikie
Note that $\gb^{\mu\nu}$ and $g^{\mu\nu}$ are treated as independent 
dynamical variables in this action. It can be easily checked that 
the Euler-Lagrange equations of $\bar{I}$ are equivalent to those of 
$I$.

In order to understand that $\bar{I}$ describes the Einstein frame, 
we rewrite it to an explicit form consisting of the Einstein-Hilbert 
action of $\gb^{\mu\nu}$ and an auxiliary matter field. The following 
general relation is useful for such a purpose: 
$ R_{\mu\nu}(g^{\al\be}, \maru_{\om} g^{\al\be},
               \maru_{\tau} \maru_{\om} g^{\al\be})
    = \bar{R}_{\mu\nu}(\gb^{\al\be}, \maru_{\om} \gb^{\al\be},
                       \maru_{\tau} \maru_{\om} \gb^{\al\be})
    + \bar{\nabla}_{\al} F^{\al}_{\mu\nu}
    - \bar{\nabla}_{\nu} F^{\al}_{\mu\al}
    + F^{\al}_{\al\be}F^{\be}_{\mu\nu}
    - F^{\al}_{\mu\al}F^{\be}_{\al\nu} $, 
where $\bar{\nabla}_{\mu}$ is the covariant derivative with respect 
to $\gb^{\mu\nu}$ and $F^{\al}_{\mu\nu}$ is defined by 
$ F^{\al}_{\mu\nu} \equiv
        \Gamma^{\al}_{\mu\nu} - \bar{\Gamma}^{\al}_{\mu\nu}
      = (1/2) \, g^{\al\be}
          ( \bar{\nabla}_{\nu} g_{\mu\be}
          + \bar{\nabla}_{\mu} g_{\be\nu}
          - \bar{\nabla}_{\be} g_{\mu\nu}) $. 
By substituting this relation into $\bar{I}$, we obtain
  \sikib
    \bar{I} =
      \frac{1}{16\pi G} \int d^3 x \, \sqrt{-\gb}
    \left[ \; \gb^{\mu\nu}
        \bar{R}_{\mu\nu}(\gb^{\al\be}, \maru_{\om} \gb^{\al\be},
                           \maru_{\tau} \maru_{\om} \gb^{\al\be})
      + \gb^{\mu\nu} ( F^{\al}_{\al\be}F^{\be}_{\mu\nu}
                     - F^{\al}_{\mu\al}F^{\be}_{\al\nu} )
    \nonumber \right. \\
    \left.
      - \gb^{\mu\nu} R_{\mu\nu}(\gb^{\al\be},g^{\al\be})
      + \frac{ \sqrt{-g} }{ \sqrt{-\gb} } \,
          f(R_{\mu\nu}(\gb^{\al\be},g^{\al\be}),g^{\mu\nu}) \;
      \right] + \mbox{surface terms} \, .
  \label{eq-FT8}
  \sikie
The Einstein frame described by $\bar{I}$ is understood as the 
system consisting of the Einstein gravity, $\gb^{\mu\nu}$, and an 
auxiliary tensor matter field, $g^{\mu\nu}$.

\subsection{Central charge in the higher curvature gravity}

We restrict our treatment hereafter to the case that the right hand 
side of the Euler-Lagrange equation (\ref{eq-FT2}) vanishes, that 
is, the spacetime is of the constant curvature. Further we assume 
that the curvature $R$ is of negative. Under such conditions, the 
BTZ black hole of the form given by eq. (\ref{eq-EG3}) can exist. 
Because the spacetime of the BTZ black hole is of constant 
curvature, the following quantity $\Om$ is constant:
  \sikib
    \Om = \frac{1}{3} \, g^{\mu\nu}
            \frac{\maru f}{\maru R_{\mu\nu}} \, .
  \label{eq-HCG1}
  \sikie
Eq. (\ref{eq-FT3}) gives the metric in the Einstein frame 
$\gb^{\mu\nu}$ to be
  \sikib
    \gb^{\mu\nu} = \Om^{-2} g^{\mu\nu}
    \quad , \quad \gb_{\mu\nu} = \Om^{2} g_{\mu\nu} \, ,
  \label{eq-HCG2}
  \sikie
where $\gb_{\mu\nu}$ is defined by 
$\gb_{\mu\al}\gb^{\al\nu} = \delta_{\mu}^{\nu}$. This is just a 
conformal transformation with constant conformal factor $\Om$. The 
calculation of the central charge in the Einstein frame can be 
carried out by relating the quantities in the Einstein frame to 
those in the original frame through the conformal transformation 
(\ref{eq-HCG2}). Note that, because the isometries of the BTZ black 
hole in both frames are the same, the asymptotic symmetry Killing 
vectors (\ref{eq-EG14}) in both frames are the same. Further by 
definitions of $\widehat{G}^{ijkl}$, $\pi_i^{\;l}$, the lapse 
function and the shift vector, we can extract their relations 
between both frames: 
$ \overline{\xi}^{\mu} = \xi^{\mu} \, , \,
  \bar{\widehat{G}}^{ijkl} = \Om^{-2} \widehat{G}^{ijkl} \, , \,
  \bar{\pi}_i^{\;l} = \Om \pi_i^{\;l} \, , \,
  \bar{N}^{\bot} = \Om N^{\bot} \, , \,
  \bar{N}^{i} = \Om^2 N^{i}
$. Substituting these results into (\ref{eq-EG18}), the central 
charge is calculated to be
  \sikib
    C = \frac{l}{2G} \, g^{\mu\nu}
          \frac{\maru f}{\maru R_{\mu\nu}} \, .
  \label{eq-HCG12}
  \sikie
Although this is the central charge calculated in the Einstein 
frame, this central charge is equivalent to that in the original 
higher curvature frame.

\subsection{Computation of the BTZ black hole entropy}

With all the preparations done above, we can proceed to the 
calculation of the BTZ black hole entropy in the higher curvature 
gravity. The BTZ black hole is also of the asymptotically $AdS_3$ 
spacetime even in the higher curvature gravity. Hence the AdS/CFT 
correspondence is available \cite{ref-CC} \cite{ref-NO1} to give the 
statistical entropy of the form (\ref{eq-EG21}). When we denote the 
BTZ black hole metric in the higher curvature gravity by the same 
notation as eq. (\ref{eq-EG3}), the mass and the angular momentum in 
the higher curvature gravity are calculated through the Noether 
charge form obtained in ref. \cite{ref-IW1} to be $\Om M$ and 
$\Om J$, respectively. Then the two Virasoro eigen values $\la$ and 
$\tilde{\la}$ in the higher curvature gravity are given by the same 
procedure as eq. (\ref{eq-EG25}),
  \sikib
    \la = \Om \, \frac{1}{2}(M+J) \quad , \quad
    \tilde{\la} = \Om \, \frac{1}{2}(M-J) \, .
  \label{eq-HCG10}
  \sikie

Finally the statistical entropy of the BTZ black hole in the higher 
curvature gravity is obtained by eqs. (\ref{eq-EG21}), 
(\ref{eq-HCG12}) and (\ref{eq-HCG10}) as:
  \sikib
    S_C = \frac{\pi}{12G} \, g^{\mu\nu}
          \frac{\maru f}{\maru R_{\mu\nu}} \,
          \left[ \, \sqrt{ 8Gl(M+J) }
                + \sqrt{ 8Gl(M-J) } \, \right]
  \label{eq-HCG13}
  \sikie
On the other hand, the geometrical entropy of the black hole can be 
calculated by eq. (\ref{eq-I1}) to be
  \sikib
    S_{IW}
      = - \frac{1}{8G} \oint_{H} d\phi \sqrt{g_{\phi\phi}} \,
          \frac{ \maru f }{ \maru R_{\mu\al} } \, g^{\nu\be} \,
          \epsilon_{\mu\nu}\epsilon_{\al\be}
      =   \frac{1}{12G} \, g^{\mu\nu}
          \frac{\maru f}{\maru R_{\mu\nu}} \,
          \oint_{r_+} d\phi \sqrt{g_{\phi\phi}} \, ,
  \label{eq-HCG14}
  \sikie
By making use of the integral in (\ref{eq-EG5}), this formula 
(\ref{eq-HCG14}) coincides with the statistical entropy $S_C$ given 
by eq. (\ref{eq-HCG13}).

\subsection{Example: $f=R +aR^2 +bR_{\mu\nu}R^{\mu\nu} -2\La$}
\label{sec-EX}

We apply the results obtained above to the case: 
$f = R + a R^2 + b R_{\mu\nu}R^{\mu\nu} - 2 \La$ ($a$, $b = const.$). 
For the constant curvature spacetime, the Euler-Lagrange equation 
(\ref{eq-FT2}) becomes
  \sikib
    (1+2aR) R_{\mu\nu} +
       \frac{1}{2}
        (R +aR^2 +bR_{\al\be}R^{\al\be} -2\La) g_{\mu\nu}
    = 0 \, .
  \label{eq-HCG17}
  \sikie
For the BTZ black hole which is the negative constant curvature 
space, $\La$ is related to the Ricci scalar through eqs. 
(\ref{eq-EG2}) and (\ref{eq-HCG17}): 
$ \La = \frac{R}{6} \, [ \, 1+(b-a)R \, ] $. The conformal 
transformation (\ref{eq-HCG2}) becomes
  \sikib
    \gb_{\mu\nu} = \Om^2 g_{\mu\nu} \quad , \quad
    \Om = 1 - \frac{12 a + 4 b}{l^2} \; ,
  \label{eq-HCG20}
  \sikie
where we have used the definition of the radius of $AdS$ spacetime: 
$l = \sqrt{-6/R}$. This gives the central charge through eq. 
(\ref{eq-HCG12}) as
  \sikib
    C = \left( 1 - \frac{12 a + 4 b}{l^2} \right) \,
        \frac{3 \, l}{2 \, G} \, ,
  \label{eq-temp1}
  \sikie
which is consistent with the results obtained in ref. 
\cite{ref-NO1}. We obtain the statistical entropy in the higher 
curvature gravity through eq. (\ref{eq-HCG13})
  \sikib
    S_C = \left( 1 - \frac{12a + 4b}{l^2} \right) \,
          \frac{\pi}{4G}
          \left[ \, \sqrt{ 8Gl(M+J) }
                + \sqrt{ 8Gl(M-J) } \, \right] \, .
  \label{eq-HCG21}
  \sikie
The geometrical entropy given by eq. (\ref{eq-I1}) is calculated to 
be
  \sikib
    S_{IW}
      = \left( 1 - \frac{12 a + 4 b}{l^2} \right) \,
        \frac{1}{4G} \oint_{r_+} d\phi \sqrt{g_{\phi\phi}} \, .
  \label{eq-temp2}
  \sikie
By making use of the integral in (\ref{eq-EG5}), this equation 
(\ref{eq-temp2}) coincides with the statistical entropy $S_C$ given 
by eq. (\ref{eq-HCG21}).


\section{Summary and Discussion}\label{sec-SD}

We have shown the statistical derivation of the BTZ black hole 
entropy in the higher curvature gravity. The resultant formula 
agrees with the one derived by the Noether charge method 
\cite{ref-IW1}. As a by-product, we have obtained the central 
charge, that is, the coefficient of the Weyl anomaly, in the higher 
curvature gravity. Although the statistical entropy is calculated in 
the higher curvature gravity, because the procedure of our 
calculation is essentially the same as the Einstein gravity, some 
issues which stemmed from the use of CFT on the boundary of $AdS_3$ 
\cite{ref-C2} remain open to be resolved. 

Provided that, instead of the higher curvature gravity, we adopt the 
BTZ black hole in diffeomorphism invariant theories of gravity which 
include symmetrized covariant derivatives of Riemann tensor like 
$R_{\mu\nu\al\be ; (\om\tau)}$, the problem which arises in applying 
our procedure to calculate the central charge is that we do not have 
the formalism of transforming the frames in such theories. However, 
because the relation (\ref{eq-EG2}) denotes that the terms of 
covariant derivative of Riemann tensor in such Lagrangians vanishes, 
it is expected that the frame transformation between the original 
and Einstein frames is also the conformal transformation with 
constant conformal factor. For the other case of Lagrangians include 
some matter fields coupling to the gravity, the factor $\Om$ should 
depend on the matter fields. However it is possible to set such 
matter fields be constant without loss of consistency with the BTZ 
black hole. Then our computation of the statistical entropy of the 
BTZ black hole in this paper is available for the general 
diffeomorphism invariant theories of gravity which include the 
symmetrized covariant derivatives of Riemann tensor and the matter 
fields coupling to the gravity, provided that we can construct a 
well-defined transformation of the frames in such theories. Thus, it 
is natural to conjecture that the equality between the geometrical 
entropy and the statistical one is retained in the general 
diffeomorphism invariant theories of gravity. Further, with noting 
the work \cite{ref-C3} which extends the computation of the 
statistical entropy of the BTZ black hole to any dimension in the 
Einstein gravity, it is expected that the equivalence between the 
geometrical and statistical entropies can be extended to any 
dimensional diffeomorphism invariant theories of gravity.

The derivation of the statistical entropy in this paper is heavily 
relied on the conformal field theory constructed on the boundary of 
the BTZ black hole spacetime. It has been well 
recognized that the unitarity of the conformal field theory requires 
the positivity of the central charge, which is consistent with eq. 
(\ref{eq-EG20}). Further it is required through eq. (\ref{eq-HCG12}) 
that $\Om>0$. On the other hand, this relation, $\Om>0$, is required 
from the null energy condition in the Einstein frame 
and the weak cosmic censorship conjecture \cite{ref-NW} in 
considering the second law of the Black hole thermodynamics for the 
geometrical entropy as mentioned below \cite{ref-JKM2}. Both of 
these conditions are the essential assumptions in establishing the 
second law in the Einstein gravity (the area theorem). For the case 
of the Lagrangian polynomial in $R$, it has already been explicitly 
derived in ref. \cite{ref-JKM2} that the null energy condition in 
the Einstein frame requires the positivity of the conformal factor 
$\Om$ of the frame transformation. This suggests that this 
requirement can be extended to the case of general higher curvature 
gravity. Further, with making use of the frame transforation in 
considering the second law in the diffeomorphism invariant theories 
of gravity, the cosmic censorship conjecture will be required to be 
satisfied in the Einstein frame. It will also be necessary to retain 
the disappearance of naked singularities in the original frame. Then 
we require the positivity of $\Om$ because, for the case that 
$\Om \leq 0$, there is a possibility of appearing singularities in 
the original frame. Thus, the cosmic censorship conjecture and the 
null energy condition in the Einstein frame must have some relation 
with the unitarity of the conformal field theory on the boundary of 
spacetime. The issue about the details of such a relation is left as 
an interesting open question.

We turn our attention to some possible applications of our results 
to some interesting cases. As we have already seen, the statistical 
black hole entropy can be computed by using the data on the boundary 
of the black hole spacetime which asymptotes to that of anti-de 
Sitter spacetime. This kind of correspondence between the symmetry 
of the boundary of $AdS$ spacetime and the conformal field theory 
(CFT) has been intensively studied recently. This correspondence can 
be used to deduce the entropy of the black string system in higher 
dimensions \cite{ref-BS}. It is interesting to investigate whether or 
not our results of this paper can be extended to the higher 
dimensional black objects. 

For the calculation of the statistical entropy through the AdS/CFT 
correspondence, the central importance exists in the correspondence 
between the symmetry of the boundary of $AdS$ and the isometry of 
the bulk spacetime of a black hole, such as the stationarity and 
axisymmetry. Hence, it is not obvious whether or not the statistical 
entropy which we have calculated can be extended to the dynamical 
cases such as the collapsing black holes and the evaporating black 
holes. As the geometrical entropy is extended with retaining its 
meaning even to the dynamical systems \cite{ref-IW1} 
\cite{ref-JKM1}, however, it is expected to be able to extend our 
results to the dynamical cases.


\section*{Acknowledgements}

We would like to Thank M.Sakagami for his useful discussions. This 
work was supported in part by Monbusho Grant-in-Aid for Scientific 
Research No.10740118.



\begin{thebibliography}{99}
\bibitem{ref-NW}
    The null energy condition is expressed as 
    $T_{\mu\nu}k^{\mu}k^{\nu} \geq 0$ for any null vector $k^{\mu}$. 
    The weak cosmic cesorship conjecture means that any naked 
    singularities never appear in the interesting dynamical 
    processes.
\end{thebibliography}
\end{document}